\documentclass{article}
\usepackage{spconf,amsmath,graphicx}
\usepackage{multirow}
\usepackage{array} 
\usepackage{threeparttable}
\usepackage{spconf,amsmath,graphicx}
\usepackage{booktabs}
\usepackage{xcolor}
\usepackage{mathrsfs}
\usepackage{hyperref}

\title{Gradient weighting for speaker verification in extremely low Signal-to-Noise Ratio}
\name{Yi Ma\textsuperscript{1}, Kong Aik Lee\textsuperscript{2}, Ville Hautam\"aki\textsuperscript{3}, Meng Ge\textsuperscript{1}, Haizhou Li\textsuperscript{4,5,1}}
\address{\textsuperscript{1}Department of Electrical and Computer Engineering, National University of Singapore, Singapore\\
\textsuperscript{2}Department of Electrical and Electronic Engineering, The Hong Kong Polytechnic University, Hong Kong\\
\textsuperscript{3} School of Computing, University of Eastern Finland, Finland\\
\textsuperscript{4}Shenzhen Research Institute of Big Data, Shenzhen, China\\
\textsuperscript{5}School of Data Science, The Chinese University of Hong Kong, Shenzhen, China}
\begin{document}
\ninept
\maketitle
\begin{abstract}
Speaker verification is hampered by background noise, particularly at extremely low Signal-to-Noise Ratio (SNR) under~0~dB. It is difficult to suppress noise without introducing unwanted artifacts, which adversely affects speaker verification.
We proposed the mechanism called Gradient Weighting (Grad-W), which dynamically identifies and reduces artifact noise during prediction. The mechanism is based on the property that the gradient indicates which parts of the input the model is paying attention to.
Specifically, when the speaker network focuses on a region in the denoised utterance but not on the clean counterpart, we consider it artifact noise and assign higher weights for this region during optimization of enhancement. 
We validate it by training an enhancement model and testing the enhanced utterance on speaker verification. 
The experimental results show that our approach effectively reduces artifact noise, improving speaker verification across various SNR levels. 
\end{abstract}

\section{Introduction}
\label{sec:intro}
Neural netowrk methods in speaker recognition and related tasks in noisy conditions have made much progress~\cite{shon2019voiceid, kataria2020feature, ma2021pl,liu2023disentangling,9747021}. However,  a low SNR condition, e.g. SNR under 0dB, still degrades the performance of speaker verification. This is due to the fact that suppressing noise without creating artifacts in human speech is challenging in low SNR environments~\cite{mamun23_interspeech}. 


Reducing noise in extremely low SNR conditions is gaining attention~\cite{mamun23_interspeech,wang2023cross,hao19_interspeech}. In the case of speaker verification in a noisy environment, one approach is to jointly train using enhancement loss and speaker classification loss~\cite{zhang2021towards,kim2022extended}. However, multiple loss functions may lead to gradient conflict ~\cite{wu21c_interspeech}. The enhancement model can also be optimized using the information used to recognize speakers~\cite{shon2019voiceid}. For example, training on the hidden activation map of the reference utterance generated from the pre-trained speaker network are proposed in~\cite{ kataria2020feature, ma2021pl,kataria2020analysis}. Although these methods are proven to be efficient, artifacts are not explicitly considered, nor is SNR below 0 dB taken into account.   To solve this problem, we train an enhancement model to filter out noise for a pre-trained and fixed speaker recognition model. The speaker model is used as both downstream network to extract speaker embedding and auxiliary model to optimize the enhancement model. Using this method, we believe it serves as a general function for other downstream tasks in a noisy environment.

The input-output behaviour of deep neural networks is characterized by the gradients generated from the logits of the target class with respect to the hidden activations or input features. The gradients alone~\cite{sundararajan2017axiomatic} or gradients multiplied with a hidden activation map~\cite{8237336,9462463} can be used to visualize the internal mechanisms of neural networks. As gradients reveal critical attributes for model predictions, they are utilized in adversarial training method~\cite{ganz2022perceptually,boopathy2020proper} and efficient learning methods~\cite{Xingyi22DeRy,Xinyin23NeurIPS,Xinyin23DeepCache,Songhua22DD}. They assume that even if input features appear visually similar, the position on which the network focuses for prediction is entirely different. In the task of speaker verification, methods based on gradients were used to visualize the working mechanism of a speaker recognition network~\cite{li22l_interspeech} and to understand the role data augmentation plays in speaker recognition~\cite{li2023visualizing}. Although gradients reflect where the network relied on to predict efficiently, to the best of our knowledge, this paper presents the first attempt to apply this property for denoising purposes. 

We leverage the gradient of a pre-trained network to detect and mitigate the artifact noise during enhancement. The highlighted regions in the gradient indicate the areas that the model is focusing on to make predictions. Based on our hypothesis, if a region is highlighted in an enhanced sentence but not in the clean counterpart, then that position corresponds to the artifact noise. 
Leveraging this observation, we build a U-Net trained with the proposed Gradient Weighting (Grad-W) to handle the noise for speaker verification in extremely low SNR conditions. The probability of artifact noise present in each time-frequency (t-f) bin is computed for each enhanced utterance using the gradient. The U-Net is optimized using the $L_1$ deviation of the activation map of the clean and enhanced features weighted with the probability of artifact noise.
To validate our method, we use noisy utterances with a wide range of SNR (from -15 dB to 15 dB) to test the speaker verification task. In each test SNR condition, we observe consistent gains in speaker verification performance. A series of ablation studies are also conducted to demonstrate how the mechanism works. The code of enhancement network training is available at: \url{https://github.com/mmmmayi/Grad-W}.

\section{Proposed method}
\subsection{Overview}
Let $R(t,f)$ be the spectrogram of the reference clean speech $r(\tau)$ 
in time-frequency domain. The speech corrupted by noise is written as $x(\tau)=r(\tau)+n(\tau)$ with the corresponding spectrogram as $X(t,f)$, where $n(\tau)$ is the noise. 
The objective of this paper is to denoise the corrupted speech to extract robust speaker embedding. The mask $M(t,f)$ generated from an enhancement model is used to denoise the noisy speech: $E(t,f)=M(t,f)\times X(t,f)$. 

The goal is to train an enhancement model for downstream speaker verification task in noisy environments. First, a residual neural network (ResNet) $\mathcal{E}$ 
is pre-trained for speaker recognition using $R(t,f)$.
After that, the weights of $\mathcal{E}$ are fixed and we train the enhancement network $\mathcal{G}$ to generate the mask $M(t,f)$ using the pair of $(R(t,f), X(t,f))$ as training data.  
To optimize $\mathcal{G}$, the gradient and activation maps generated from the fixed $\mathcal{E}$ is used as a part of the loss function. The enhanced utterance $E(t,f)$ is fed into $\mathcal{E}$ to generate robust speaker embedding.

\begin{figure}[!h]
\centering
\includegraphics[trim=0.7cm 0.7cm 0.5cm 0.7cm, clip,width=8cm]{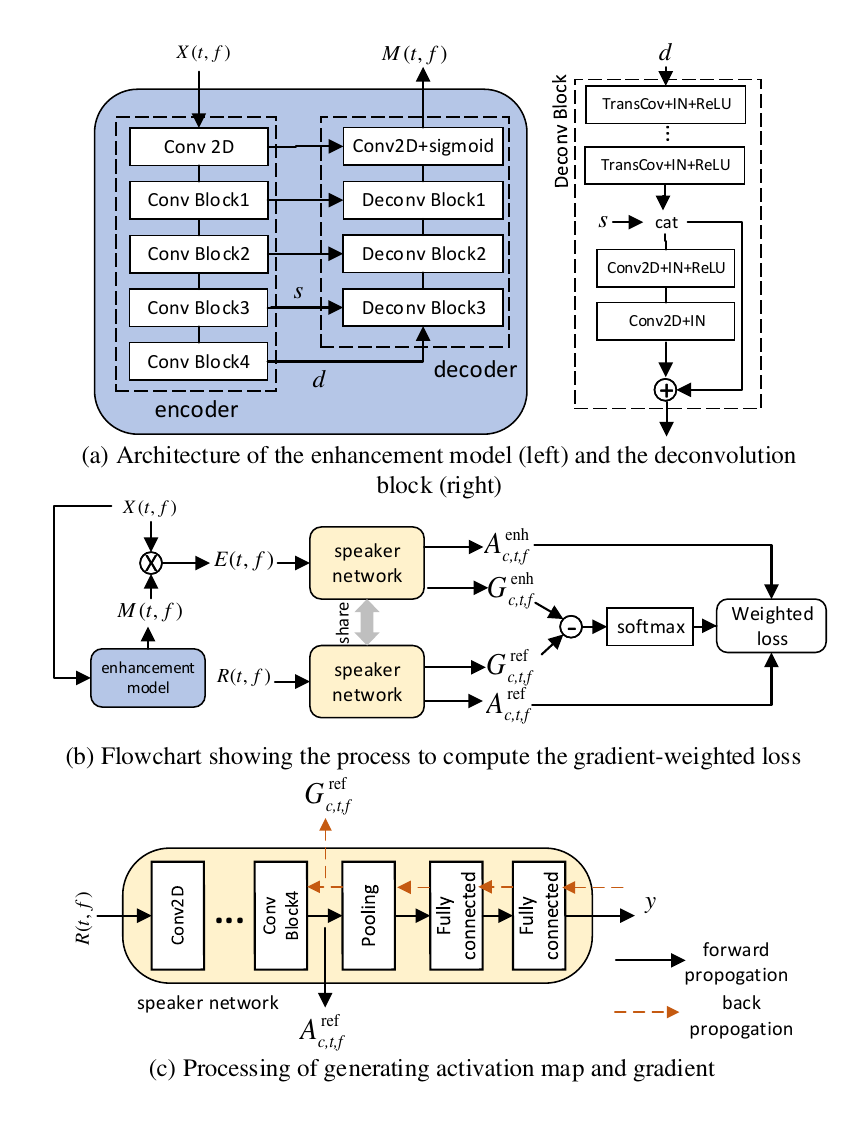}

\caption{To optimize the enhancement model (a), we generate activation maps and gradients from a pre-trained and fixed speaker network in (c). The activation map and gradients are used in computing the loss in (b). }
\label{framework}
\vspace{-0.5cm}
\end{figure}
\subsection{U-Network enhancement model}
As shown in Fig.~\ref{framework}(a), our enhancement network consists of two parts: an encoder and decoder, connected by skip connections.The encoder architecture is the same with the frame-level layers of $\mathcal{E}$. The first layer is a convolution layer with a channel dimension of 32, followed by a batch normalization layer and the ReLU activation function. This is followed by four convolutional blocks (Conv Block), with 6, 8, 12, and 6 convolution layers in each block. These four convolution blocks, each downsamples the input's time and frequency dimensions by a factor of two, while the channel dimension is upsampled by a factor of two.  

We adopt the mask prediction network proposed in~\cite{dabkowski2017real} as the decoder. The decoder consists of three deconvolution blocks (Deconv Block) and one convolution layer. Each Deconv Block contains 6, 4, and 3 transposed convolution operators.  The output of the last Deconv Block $d$ (output of encoder for the last Deconv Block), as well as the output from each corresponding encoder block $s$ is used as input of each Deconv Block. The former is processed using transposed convolution operators (TransCov) with instance normalization (IN) and ReLU. Then the output is concatenated with $s$ and two convolution layers are applied with residual connection. In the final step, the mask is generated using one convolution layer and Sigmoid as the activation function.

\subsection{Artifact noise detection}
Gradients reflect how the speaker network recognizes speakers, and we consider them to be auxiliary information when optimizing the enhancement model to adapt to downstream speaker verification task. We propose to use the gradient to locate the artifact noise generated in enhancement utterances. 

Specifically, the gradient used in this paper is derived from the target speaker logit $y$ with respect to the hidden activation map of the pre-trained $\mathcal{E}$. The process for clean utterance as an example is shown in Fig.~\ref{framework}(c). Formally, it is written as:
\begin{equation}
\vspace{-0.3cm}
G^\text{ref}_{c,t,f}=\frac{\partial y}{\partial A^\text{ref}_{c,t,f}},
\end{equation}
where $A^\text{ref}_{c,t,f}$ is the output of the layer before pooling: 
\begin{equation}
A^\text{ref}_{c,t,f}=\mathcal{E}(R(t,f)).
\end{equation}
Here $c$, $t$, and $f$ denote the index of channel, time and frequency. Both $A^\text{ref}_{c,t,f}$ and $G^\text{ref}_{c,t,f}$ are matrices of size \textit{channel}$\times$ \textit{time}$ \times$\textit{frequency}. 
Using the gradient generated from the clean utterance, we can determine which parts of the hidden activation map are likely to contain information about the target speaker. 
Gradients with higher values indicate that the model utilizes more speaker information at the corresponding position when making predictions.
Similarly,  $A^\text{enh}_{c,t,f}$ and $G^\text{enh}_{c,t,f}$ are computed for enhanced utterance $E(t,f)$.

In~\cite{li2023visualizing}, it was shown how speaker recognition networks, trained through data augmentation, handle noise: the model reduces its reliance on extracting speaker information from segments corrupted by noise. We hypothesise that clean gradient $G^\text{ref}_{c,t,f}$ serves as the upper bound for $G^\text{enh}_{c,t,f}$ considering that $G^\text{enh}_{c,t,f}$ is derived from the enhanced utterance. 
If artifact noise appears, $G^\text{enh}_{c,t,f}$ in the corresponding position would be larger than $G^\text{ref}_{c,t,f}$ since clean utterances contain little speaker information at the artifact noise's position. 
As shown in shown in Fig.~\ref{framework}(b), we compute the distance $D_{t,f}$ between $G^\text{enh}_{c,t,f}$ and $G^\text{ref}_{c,t,f}$ in each time-frequency (t-f) bin by summing the distance across the channels first:
\begin{equation}
\label{distance}
D_{t,f}=\sum_c(G^\text{enh}_{c,t,f}-G^\text{ref}_{c,t,f}).
\vspace{-0.2cm}
\end{equation}
The resulting distance is used to compute the probability $P_{t,f}$ of the presence of artifact noise in each time-frequency bin via a softmax function:
\begin{equation}
\label{probability}
P_{t,f}=\frac{\exp(D_{t,f})}{\sum_{t,f}\exp(D_{t,f})}.
\vspace{-0.5cm}
\end{equation}


\subsection{Gradient weighting}
\label{weighting loss}
To reduce the artifact noise produced during enhancement, we compute the $L_1$ deviation of the activation map of clean and enhanced signal with $P_{t,f}$ as the weight for each t-f bin:
\begin{equation}
\label{w-dfl}
\vspace{-0.3cm}
\mathcal{L}_\text{Grad-W}=\sum_{c}\sum_{t,f}\lvert A^\text{ref}_{c,t,f}-A^\text{enh}_{c,t,f}  \rvert\times P_{t,f}.
\end{equation}
The higher the probability of artifact noise occurrence, the greater the penalty applied to the corresponding position.
We only retain the probability $P_{t,f}$ within the time-frequency domain to ensure training stability. During optimization, the $P_{t,f}$ is applied on each channel of the $L_1$ deviation. 

We note that if all t-f bins are assigned equal weights as one, the loss function in~\eqref{w-dfl} reduces to:
\begin{equation}
\label{dfl}
\mathcal{L}_\text{Equal-W}=\sum_{c,t,f}\lvert A^\text{ref}_{c,t,f}-A^\text{enh}_{c,t,f}  \rvert.
\vspace{-0.3cm}
\end{equation}
This is one of the baselines we use to determine whether or not the weight proposed is efficient.
We note that the~\eqref{dfl} is similar with \textit{deep feature loss}~\cite{kataria2020feature}, while the activation of all layers is used instead of just the layer before the pooling layer to calculate the deviation. We will discuss this difference in Section~\ref{results}.

\section{Experiment}
\subsection{Dataset and metrics}
The train set of VoxCeleb2~\cite{chung2018voxceleb2} is used to train the ResNet-34 as speaker recognition network. The training data is augmented with MUSAN dataset (speech, noise, music)~\cite{snyder2015musan} and the RIR dataset (reverb)~\cite{kim2017generation}. There is a 40\%, 30\%, and 30\% ratio between the samples without data augmentation and the samples with data augmentation from the MUSAN dataset or the RIR dataset, respectively. A random selection is made from [0, 5, 10, 15] and [5, 8, 10, 15] to determine the SNR for utterances augmented with noise and music. Three to seven samples from speech class are used to generate a babble, and the SNR for each sample is calculated using [13, 15, 17, 20].

To generate simulated noisy utterances, training data of VoxCeleb2 is corrupted with MUSAN dataset during enhancement model training. The simulated noisy utterances are only used for enhancement model training. There is a random selection of SNRs between -10dB and 0dB for the noise and music classes in MUSAN. The babble noise is formed by combining 5 to 8 utterances in speech class of MUSAN, with SNR for each selected from $[5, 8, 10, 13, 15]$. 100 speakers in VoxCeleb2 are used as validation set. 

Using Vox1-O, we designed a series of simulated noisy test sets. The noise drawn from DNS-challenge~\cite{reddy2020interspeech} corrupts Vox1-O and results in seven test SNRs (-15dB, -10dB, -5dB, 0dB, 5dB, 10dB, 15dB). The original Vox1-O is used to assess the enhancement impact on clean speech as well. The Equal Error Rate (EER) and minimum Detection Cost Function (minDCF) are used to evaluate our system.

\subsection{Implementation details}
Our speaker recognition network is implemented using WeSpeaker toolkit~\cite{wang2022wespeaker}. 
Adam optimizer is used to train the enhancement network with the learning rate as 0.0005 (warmup for 5 epochs). We set the batch size as 32 and trained it for 50 epoch. We use Fbank to extract the time-frequency domain features for both speaker and enhancement network in this paper.

\section{Results}
\label{results}

  \begin{table*}[h]
  \vspace{-0.5cm}
  \fontsize{9}{10}\selectfont
    \caption{Results of the baseline and our proposed models.  The text in bold is the best performance  for each metric in different SNRs. ``Noisy'' denotes no enhancement process before speaker network; ``Equal-W'' column indicates no weight assigned to $L_1 $ deviation.}
  \label{baseline}
  \centering
  \setlength\tabcolsep{3pt}
  \begin{tabular}{c|cc|cc|cc|cc|cc}
  \toprule
    & \multicolumn{2}{c|}{Noisy} & \multicolumn{2}{c|}{VoiceID~\cite{shon2019voiceid}}     & \multicolumn{2}{c|}{DFL~\cite{kataria2020feature}} & \multicolumn{2}{c|}{Equal-W}           & \multicolumn{2}{c}{Grad-W (Ours)}           \\ \hline
SNR(dB) & \multicolumn{1}{c}{EER}  & minDCF & \multicolumn{1}{c}{EER} & minDCF & \multicolumn{1}{c}{EER}    & minDCF   & \multicolumn{1}{c}{EER} &minDCF & \multicolumn{1}{c}{EER} & minDCF \\
   \hline
  -15  & \multicolumn{1}{c}{33.52}&1.0000& \multicolumn{1}{c}{26.73}&0.9998& \multicolumn{1}{c}{27.8}&\textbf{0.9991}& \multicolumn{1}{c}{22.73}&0.9996& \multicolumn{1}{c}{\textbf{21.40}}&0.9999\\
-10 & \multicolumn{1}{c}{18.70}&0.9890& \multicolumn{1}{c}{14.67}&0.6747& \multicolumn{1}{c}{19.64}&0.9819& \multicolumn{1}{c}{11.44}&0.5979& \multicolumn{1}{c}{\textbf{10.89}}&\textbf{0.5670}\\
  -5  & \multicolumn{1}{c}{7.87}&0.4344& \multicolumn{1}{c}{7.27}&0.3888& \multicolumn{1}{c}{13.66}&0.7982& \multicolumn{1}{c}{6.01}&0.3559& \multicolumn{1}{c}{\textbf{5.65}}&\textbf{0.3443}\\
0  & \multicolumn{1}{c}{4.47}&0.2832& \multicolumn{1}{c}{4.41}&0.2742& \multicolumn{1}{c}{7.90}&0.5067& \multicolumn{1}{c}{3.92}&0.2469& \multicolumn{1}{c}{\textbf{3.91}}&\textbf{0.2375}\\
5  & \multicolumn{1}{c}{3.01}&0.1818& \multicolumn{1}{c}{3.00}&0.1900& \multicolumn{1}{c}{4.16}&0.2668& \multicolumn{1}{c}{2.76}&0.1752& \multicolumn{1}{c}{\textbf{2.73}}&\textbf{0.1734}\\
10  & \multicolumn{1}{c}{2.20}&0.1447& \multicolumn{1}{c}{2.34}&0.1412& \multicolumn{1}{c}{2.69}&0.1705& \multicolumn{1}{c}{2.09}&0.1403& \multicolumn{1}{c}{\textbf{2.06}}&\textbf{0.1344}\\
15  & \multicolumn{1}{c}{1.83}&0.1042& \multicolumn{1}{c}{1.88}&0.1116& \multicolumn{1}{c}{1.92}&0.1175& \multicolumn{1}{c}{1.75}&0.0996& \multicolumn{1}{c}{\textbf{1.73}}&\textbf{0.0979}\\

clean  & \multicolumn{1}{c}{1.24}&\textbf{0.0768}& \multicolumn{1}{c}{1.30}&0.0787 & \multicolumn{1}{c}{1.25}&0.0783& \multicolumn{1}{c}{1.24}&0.0769& \multicolumn{1}{c}{\textbf{1.22}}&0.0782\\\hline
Average  & \multicolumn{1}{c}{9.11}&0.4018& \multicolumn{1}{c}{7.70}&0.3574& \multicolumn{1}{c}{9.88}&0.4899& \multicolumn{1}{c}{6.49}&0.3365& \multicolumn{1}{c}{\textbf{6.20}}&\textbf{0.3291}        \\
\bottomrule

\end{tabular}


\end{table*}

\label{results}
  \begin{table*}[h]
  \begin{threeparttable}
  \fontsize{9}{10}\selectfont
    \caption{Results of the ablation study. ``Clean-W'' represents the weights generated using clean utterance; ``Res-W'' denotes the gradient weighting used in a residual way. ``No softmax'' represents replacing softmax with min-max normlization; ``Channel'' indicates the distance computed in channel domain; ``Residual'' denotes the weight is computed in terms of residual noise and ``Both'' for both residual and artifact noise. The row of $D$ and $P$ show the computation of distance and probability of each experiment. We sum $D$ along the channel domain unless indicated otherwise. The symbol $\sum_c$ is omitted for brevity.}
  \label{ablation}
  \centering
  \setlength\tabcolsep{3pt}
  \begin{tabular}{c|cc|cc|cc|cc|cc|cc|cc}
  \toprule
    & \multicolumn{2}{c|}{Grad-W (Ours)} & \multicolumn{2}{c|}{Clean-W}     & \multicolumn{2}{c|}{Res-W} & \multicolumn{2}{c|}{No softmax}         & \multicolumn{2}{c|}{Channel\tnote{*}}& \multicolumn{2}{c|}{Residual} &\multicolumn{2}{c}{Both} \\ \hline
    $D$&\multicolumn{2}{c|}{$G^\text{enh}_{c,t,f}-G^\text{ref}_{c,t,f}$}&\multicolumn{2}{c|}{$G^\text{ref}_{c,t,f}$}&\multicolumn{2}{c|}{$G^\text{enh}_{c,t,f}-G^\text{ref}_{c,t,f}$}&\multicolumn{2}{c|}{$G^\text{enh}_{c,t,f}-G^\text{ref}_{c,t,f}$}&\multicolumn{2}{c|}{$G^\text{enh}_{c,t,f}-G^\text{ref}_{c,t,f}$}&\multicolumn{2}{c|}{$G^\text{ref}_{c,t,f}-G^\text{enh}_{c,t,f}$}&\multicolumn{2}{c}{$|G^\text{enh}_{c,t,f}-G^\text{ref}_{c,t,f}|$}\\\hline
     $P$&\multicolumn{2}{c|}{softmax}&\multicolumn{2}{c|}{softmax}&\multicolumn{2}{c|}{softmax+1}&\multicolumn{2}{c|}{min-max norm}&\multicolumn{2}{c|}{softmax}&\multicolumn{2}{c|}{softmax}&\multicolumn{2}{c}{softmax}\\\hline
SNR(dB) & \multicolumn{1}{c}{EER}  & minDCF & \multicolumn{1}{c}{EER} & minDCF & \multicolumn{1}{c}{EER}    & minDCF   & \multicolumn{1}{c}{EER} &minDCF & \multicolumn{1}{c}{EER} & minDCF& \multicolumn{1}{c}{EER} & minDCF& \multicolumn{1}{c}{EER} & minDCF \\
   \hline
 -15  & \multicolumn{1}{c}{\textbf{21.40}}&0.9999& \multicolumn{1}{c}{22.18}&1.0000& \multicolumn{1}{c}{22.16}&0.9998& \multicolumn{1}{c}{22.14}&0.9998& \multicolumn{1}{c}{22.04}&\textbf{0.9732}& \multicolumn{1}{c}{22.43}&1.0000& \multicolumn{1}{c}{23.12}&0.9999\\
-10  & \multicolumn{1}{c}{\textbf{10.89}}&0.5670& \multicolumn{1}{c}{11.16}&0.5727& \multicolumn{1}{c}{11.27}&0.5983& \multicolumn{1}{c}{11.01}&0.5729& \multicolumn{1}{c}{11.81}&\textbf{0.5536}& \multicolumn{1}{c}{11.45}&0.5914& \multicolumn{1}{c}{11.70}&0.6178\\
 -5  & \multicolumn{1}{c}{\textbf{5.65}}&\textbf{0.3443}& \multicolumn{1}{c}{5.73}&0.3479& \multicolumn{1}{c}{5.77}&0.3575& \multicolumn{1}{c}{5.84}&0.3512& \multicolumn{1}{c}{6.42}&0.3533& \multicolumn{1}{c}{5.78}&0.3632& \multicolumn{1}{c}{6.06}&0.3543\\
0 & \multicolumn{1}{c}{3.91}&0.2375& \multicolumn{1}{c}{\textbf{3.77}}&0.2356& \multicolumn{1}{c}{3.81}&0.2424& \multicolumn{1}{c}{3.79}&0.2448& \multicolumn{1}{c}{3.98}&0.2489& \multicolumn{1}{c}{3.79}&\textbf{0.2342}& \multicolumn{1}{c}{3.91}&0.2444\\
5  & \multicolumn{1}{c}{2.73}&0.1734& \multicolumn{1}{c}{2.75}&0.1690& \multicolumn{1}{c}{\textbf{2.72}}&0.1738& \multicolumn{1}{c}{2.75}&0.1753& \multicolumn{1}{c}{2.88}&\textbf{0.1665}& \multicolumn{1}{c}{2.73}&0.1736& \multicolumn{1}{c}{2.74}&0.1756\\
10  & \multicolumn{1}{c}{2.06}&\textbf{0.1344}& \multicolumn{1}{c}{\textbf{2.03}}&0.1360& \multicolumn{1}{c}{2.12}&0.1400& \multicolumn{1}{c}{2.08}&0.1393& \multicolumn{1}{c}{2.17}&0.1406& \multicolumn{1}{c}{2.10}&0.1365& \multicolumn{1}{c}{2.13}&0.1417\\
15  & \multicolumn{1}{c}{1.73}&0.0979& \multicolumn{1}{c}{\textbf{1.72}}&\textbf{0.0966}& \multicolumn{1}{c}{\textbf{1.72}}&0.1006& \multicolumn{1}{c}{1.75}&0.0989& \multicolumn{1}{c}{1.75}&0.1067& \multicolumn{1}{c}{1.77}&0.1026& \multicolumn{1}{c}{1.73}&0.1024\\
clean  & \multicolumn{1}{c}{\textbf{1.22}}&0.0782& \multicolumn{1}{c}{1.24}&0.0767& \multicolumn{1}{c}{1.23}&\textbf{0.0763}& \multicolumn{1}{c}{\textbf{1.22}}&0.0771& \multicolumn{1}{c}{1.24}&0.0773& \multicolumn{1}{c}{\textbf{1.22}}&0.0768& \multicolumn{1}{c}{\textbf{1.22}}&0.0764\\ \hline
Average& \multicolumn{1}{c}{\textbf{6.20}}&0.3291& \multicolumn{1}{c}{6.32}&0.3293& \multicolumn{1}{c}{6.35}&0.3361& \multicolumn{1}{c}{6.32}&0.3324& \multicolumn{1}{c}{6.54}&\textbf{0.3275}&\multicolumn{1}{c}{6.41}&0.3348&\multicolumn{1}{c}{6.58}&0.3391\\
\bottomrule

\end{tabular}
    \begin{tablenotes}
     \item[*] $D$ is computed along time-frequency domain: $D_{c}=\sum_{t,f}(G^\text{enh}_{c,t,f}-G^\text{ref}_{c,t,f})$.
    
  \end{tablenotes}


  \end{threeparttable}
  \vspace{-0.5cm}
\end{table*}

\begin{figure}[!t]
\centering
\includegraphics[trim=0.5cm 0.8cm 0cm 0.5cm, clip, width=7.5cm]{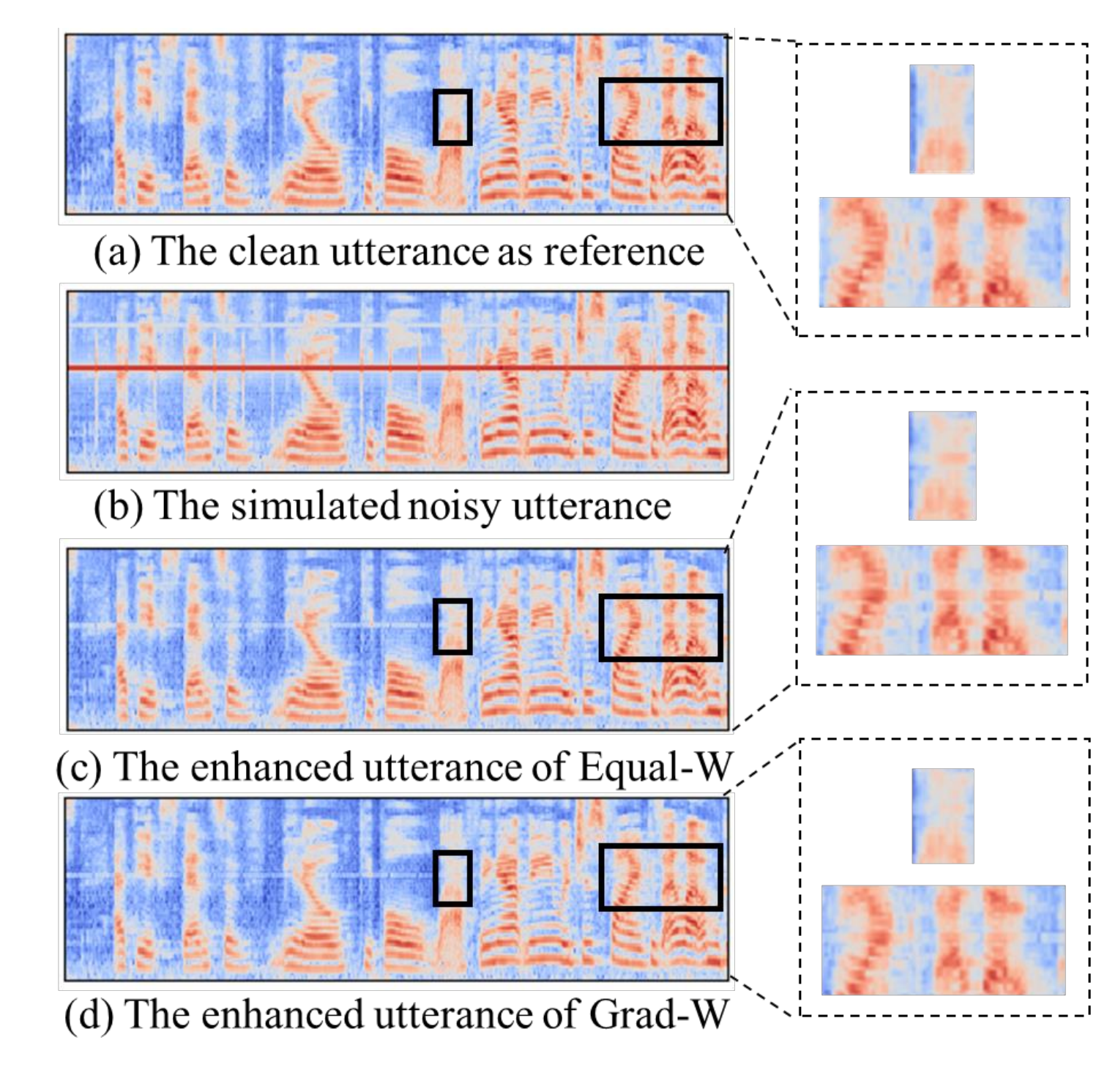}
\caption{Comparison of enhanced utterance generated from Equal-W and Grad-W. Figure best viewed in color.}
\label{enhanced result}
\vspace{-0.5cm}
\end{figure}
\subsection{Comparison with baseline system}

As shown in Table \ref{baseline}, speaker recognition system without enhancement processing (``Noisy'') is used as one of our baselines. We also show the performance of the system without assigning different weights (``Equal-W'') to validate the efficiency of gradient weighting. While VoiceID~\cite{shon2019voiceid} and Deep Feature Loss (DFL) ~\cite{kataria2020feature} are effective for speaker recognition in noisy environments, these methods have not been broadly investigated for SNRs below 0 dB. With the same training settings and data employed in our experiment, we demonstrated the performance of training our system with these two loss functions in Table \ref{baseline}. For all experiments, we show the performance of speaker verification test under both extremely low SNR (-15 to -5 dB) and regular SNR (0 to 15 dB and clean). All evaluation conditions were averaged for convenience of obeservation.

For average results, our proposed system improved EER and minDCF by 31.94\% and 18.09\%, respectively, over baseline ``Noisy''. Meanwhile, our proposed method is beneficial for speaker recognition networks under different SNRs evidently. Compared to ``Equal-W'', we find the Grad-W mechanism obtains better performance on average (6.20\% vs. 6.49\% for EER and 0.3291\% vs. 0.3365\% for minDCF), which is more significant for SNRs below 0 dB than for regular SNRs. 
In terms of existing enhancement methods ~\cite{shon2019voiceid} and ~\cite{kataria2020feature}, VoiceID still improves the baseline network efficiently, while DFL slightly degrades it. Considering the difference between DFL and ``Equal-W'', we find that using the last convolutional layer to generate the activation map is more efficient than using all layers. This may be due to the fact that deep layer activation contains information that is tied to the speaker, whereas shallow layer activation is severely corrupted. It is more difficult to optimize the shallow layers in a negative SNR condition.

\subsection{Abalation study for different settings}
In Table~\ref{ablation}, ``Clean-W'' represents gradients derived only from clean utterances to calculate $P_{t,f}$. The results show that there is still some improvement over ``Equal-W''. The reason for this is that $P_{t,f}$ here indicates the likelihood that t-f bins contain target speaker's information. 
``Res-W'' shares the same setting with our proposed function except we add 1 to the probability in $\mathcal{L}_\text{Grad-W}$: $\sum_{c}\sum_{t,f}\lvert A^\text{ref}_{c,t,f}-A^\text{enh}_{c,t,f}  \rvert\times (1+P_{t,f})$. Performance degrades slightly in comparison with ``Grad-W'', indicating that it's better to focus on only important positions in the activation map during optimization.

Next, we investigate the effect of softmax. We use min-max normalization (``No softmax'' in Table~\ref{ablation}) to convert $D_{t,f}$ to $P_{t,f}$. The results show that probability is a better way of weighting, however, the distance $D_{t,f}$ itself indicates the presence of artifact noise as well. We also assume that weight after softmax increases the sparsity of the activation map, thus reducing overfitting.

In the proposed method, we sum the distance $D$ along the channel to obtain weight in the time-frequency domain. To investigate if artifact noise can be detected in the channel domain, we sum the distance along time and frequency to generate a weight for each channel. The weight after softmax is multiplied with the $L_1$ deviation of activation map. ``Channel'' in Table~\ref{ablation} shows the performance for this setting. Although occasional metrics show the effectiveness of this approach, its performance cannot be generalized to other situations.

We compared the performance of using $\sum_c G^r_{c,t,f}-G^e_{c,t,f}$ and $\sum_c|G^r_{c,t,f}-G^e_{c,t,f}|$ as distance. As analyzed, the former aims to detect the missing speaker information in enhanced utterances, which may due to residual noise covering it. Similarly, the latter considers both the artifact and residual noise. The performance is shown as ``Residual'' and ``Both'' in Table~\ref{ablation}. Comparing with ``Equal-W'' in Table~\ref{results}, we find that separately considering artifacts (``Grad-W'') or residual noise (``Residual'') both benefit the speaker network. However, reducing them simultaneously results in performance degrades. We assume that this observation indicates the essence of weighting is to reduce overfitting by assigning different weight to activation map's element. 

We also provide some qualitative analysis in Fig.~\ref{enhanced result}. On the right side, we enlarge segments containing speaker information with black boxes. It is evident that the speaker's voice with Equal-W is distorted, which may result in degraded speaker recognition performance. However, our method clearly reduces this distortion.

\section{Conclusion}
Motivated by the discrepancy in the gradient between the clean and enhanced utterance, we have proposed a novel method to detect the artifact noise and assign a high weight to the corresponding position. Based on the proposed method, the enhancement model was trained to suppress noise in extremely low SNR for speaker verification. Experimental results show the proposed method provides benefits for robust speaker verification in a wide range of SNR. In future works, we will explore leveraging gradients in different ways, such as reducing the conflict optimization directions for multiple losses and using the gradients as reference signals in speaker extraction.

\section{Acknowledge}
This research / project is supported by Human Robot Collaborative AI under its AME Programmatic Funding Scheme (Project No. A18A2b0046); The National Natural Science Foundation of China (Grant No. 62271432); Shenzhen Science and Technology Research Fund (Fundamental Research Key Project Grant No. JCYJ20220818103001002);The Internal Project Fund from Shenzhen Research Institute of Big Data under Grant No. T00120220002; And Research Council Finland project number \#350093.


\vfill\pagebreak

\bibliographystyle{IEEEbib}
\bibliography{refs}

\end{document}